\documentstyle[11pt,psfig]{article}
\def\ifmath#1{\relax\ifmmode #1\else $#1$\fi}

\def\3quarter{{\textstyle{3 \over 4}}}

\def\ra{\rightarrow}

\def\lf{\leaders\hbox to 1em{\hss.\hss}\hfill}

\def\21{$SU(2) \ot U(1)$}

\def\VEV#1{\left\langle #1\right\rangle}

\def\lsim{\raise0.3ex\hbox{$\;<$\kern-0.75em\raise-1.1ex\hbox{$\sim\;$}}}
\def\gsim{\raise0.3ex\hbox{$\;>$\kern-0.75em\raise-1.1ex\hbox{$\sim\;$}}}

\def\bel{\begin{letter}}
\def\eel{\end{letter}}
\def\beq{\begin{equation}}
\def\eeq{\end{equation}}
\def\bef{\begin{figure}}
\def\eef{\end{figure}}
\def\bet{\begin{table}}
\def\eet{\end{table}}
\def\bea{\begin{eqnarray}}
\def\ba{\begin{array}}
\def\ea{\end{array}}
\def\bi{\begin{itemize}}
\def\ei{\end{itemize}}
\def\ben{\begin{enumerate}}
\def\een{\end{enumerate}}
\def\ra{\rightarrow}
\def\ot{\otimes}

\def\eea{\end{eqnarray}}
\def\np#1#2#3{           {\it Nucl. Phys. }{\bf #1} (19#2) #3}
\def\pl#1#2#3{           {\it Phys. Lett. }{\bf #1} (19#2) #3}
\def\pr#1#2#3{           {\it Phys. Rev. }{\bf #1} (19#2) #3}
\def\prep#1#2#3{         {\it Phys. Rep. }{\bf #1} (19#2) #3}
\def\prl#1#2#3{          {\it Phys. Rev. Lett. }{\bf #1} (19#2) #3}

\def\zp#1#2#3{           {\it Zeit. fur Physik }{\bf #1} (19#2) #3}

\def\n.c.#1#2#3{         {\it Nuovo Cim. }{\bf #1} (19#2) #3}
\def\r.n.c.#1#2#3{       {\it Riv. del Nuovo Cim. }{\bf #1} (19#2) #3}

\begin{document}
\rightline{FTUV/94-28}
\rightline{IFIC/94-26}
\begin{center}
{\bf UPDATED LIMITS ON VISIBLY AND INVISIBLY DECAYING HIGGS
BOSONS FROM LEP }\\
\vskip 0.4cm
{\bf \underline{ F. de Campos}}
\footnote{Bitnet CAMPOSC@vm.ci.uv.es - Decnet 16444::CAMPOSC}
{\bf and J. W. F. Valle}
\footnote{Bitnet VALLE@vm.ci.uv.es - Decnet 16444::VALLE}\\
Instituto de Fisica Corpuscular - IFIC/CSIC\\
Dept. de F\'isica Te\`orica, Universitat de Val\`encia\\
46100 Burjassot, Val\`encia, SPAIN\\
\vskip 0.4cm
{\bf A. Lopez-Fernandez}
\footnote{Bitnet ALFON@CERNVM}\\
PPE Division, CERN\\
CH-1211 Geneve 23, Switzerland\\
\vskip 0.2cm
and \\
\vskip 0.2cm
{\bf J. C. Rom\~ao}
\footnote{Bitnet ROMAO@PTIFM}\\
{Departamento de F\'{\i}sica, Instituto Superior T\'ecnico\\
Av. Rovisco Pais, 1 - 1096 Lisboa Codex, PORTUGAL}\\
\vskip .3cm
{\em Talk presented at the XXIX Rencontres de Moriond,\\
 Electroweak Interactions and Unified Theories, \\
March 12-19, 1994, Meribel - Savoy - FRANCE}\\
\end{center}
\vskip 0.4cm
The problem of mass generation remains one
of the main puzzles in particle physics today.
In the standard model all masses arise as a result
of the spontaneous breaking of SU(2)$\otimes$U(1) the gauge symmetry.
This implies the existence of an elementary Higgs boson
not yet found. Recently the LEP
experiments on  $e^+ e^-$ collisions around the
Z peak have placed important restrictions on
the Higgs boson mass
$
\label{1}
m_{H_{SM}} \gsim 60 \rm{GeV}.
$

There are many reasons to think that there may exist
additional Higgs bosons in nature. One such extension
of the minimal standard model is provided by supersymmetry and
the desire to tackle the hierarchy problem \cite{revsusy}.
Another reason is neutrino physics. Indeed, there
are many extensions of the minimal standard model which
induce neutrino masses either at the tree level or radiatively
through an enlargement in the Higgs sector \cite{fae}.

In many of these extensions one has the possibility
that the Higgs boson may decay into invisible particles,
such as $H \ra \chi \chi$ where $\chi$ is the lightest
neutralino in supersymmetry, possible when $2 m_\chi < M_H$.

Amongst the extensions of the standard model which have been
suggested to generate neutrino masses, the majoron
models are particularly interesting and have been
widely discussed \cite{fae}. The majoron is a Goldstone
boson associated with the spontaneous breaking of the lepton
number. Astrophysical arguments based from stellar cooling rates
constrain its couplings to the charged fermions \cite{KIM},
while the LEP measurements of the invisible Z width restrict
the majoron couplings to the gauge bosons in an important
way. In particular, models where the majoron is not a
singlet \cite{GR} under the SU(2)$\otimes$U(1) symmetry are
now excluded \cite{LEP1}.

There is, however, a wide class of models
\cite{JoshipuraValle92}, motivated by neutrino physics,
which are characterized by the spontaneous violation
of a global $U(1)$ lepton number symmetry by an SU(2)$\otimes$U(1)
singlet vacuum expectation value $\VEV{\sigma}$ \cite{CMP}.
These models may naturally explain the neutrino masses required
by astrophysical and cosmological observations \cite{BGLAST}.

Another example is provided by supersymmetric
extensions of the standard model where R parity is
spontaneously violated \cite{MASI}.

In all these extensions of the minimal standard model
the global $U(1)$ lepton number symmetry is spontaneously
violated close to the electroweak scale.
Such a low scale for the lepton number violation
is preferred since, in these models, $m_\nu \to 0$
as $\VEV{\sigma} \ra 0$.
As a result, a relatively low value of $\VEV{\sigma}$ is
required in order to obtain naturally small neutrino
masses \cite{JoshipuraValle92}.

In any model with a spontaneous violation of a
global $U(1)$ symmetry around the weak scale
(or below) the corresponding Goldstone boson
has significant couplings to the Higgs bosons,
even if its other couplings are suppressed.
This implies that the Higgs boson can decay
with a substantial branching ratio into the
invisible mode \cite{JoshipuraValle92,Joshi92,HJJ}
$h \ra J\;+\;J
\label{JJ}$ where $J$ denotes the majoron.

Such an invisible Higgs decay would lead to events
with large missing energy that could be observable
at LEP and affect the corresponding Higgs mass bounds.
In order to do this we determine the Higgs
boson production and visible and invisible decay rates,
which involves three independent
parameters: the Higgs boson mass $M_H$, its coupling
strength to the Z, normalized by that of the standard model, we call
this factor $\epsilon^2$, and the invisible Higgs boson
decay branching ratio.

We have used the same method as described in \cite{Alfon},
in order to deduce the regions in the parameter
space of the model that can be ruled out already.
The procedure was the following:
In each case we have allowed the Higgs boson to decay
into invisible channels with an arbitrary branching ratio
between 0 and 1, and kept the weakest limit. In order
to do this we have first considered the two extreme
limits where the Higgs boson decays 100\% visibly or
invisibly; we obtained the absolute bound on the
coupling $ZZh$ ($\epsilon^2$) as a function of $m_H$.
The analysis was made considering the two differents
signals given in the Higgs boson decay: the invisible case and
the visible one.

For the invisible case, namely:
$Z \ra H Z^*$,$H \ra $invisible, $Z \ra q\bar{q}$,
we used directly the results presented by
the Aleph Collaboration in reference \cite{ALEPH}, in which they
set a limit on the maximum allowed coupling of the Higgs to the Z$^0$
as a function of its mass.

For the visible case:
$Z \ra H Z^*$, $H \ra q\bar{q}$, $Z \ra \nu \nu$ or $ll$
where we directly applied the limits from the standard
searches, using results from all experiments
\cite{Delphi,ALEPH,L3,OPAL}. In these case the background
events were considered where they existed. For all
values of the Higgs mass, we found that the
weakest limit was obtained in the case of 100\% standard decays,
since in this channel it is not possible to tag the events in which
the Z$^0$ decays hadronically due to the large background from normal
processes.

As an illustration we show in Figure 1 the exclusion contours
in the plane $\epsilon^2$
vs. $BR(H \ra $ visible) for the particular choice for the Higgs
mass $M_H = 50$ GeV. The two curves corresponding to the searches
for visible and invisible decays are combined to give the final bound;
values of $\epsilon^2$ above 0.1 are ruled out independently of the
value of $BR(H \ra $ visible). The solid line in Figure 2 shows
the region in the $\epsilon^2$ vs. $M_H$ that can be excluded by
the present LEP analyses, independent of the mode of Higgs decay,
visible or invisible. An analysis on
invisible higgs bosons can also be found in \cite{DP}.

We have also estimated the additional range of parameters that can
be covered by LEPII. We assumed that the total luminosity collected
will be 500 pb$^{-1}$, and give the results for two values of the
centre-of-mass energy: 175 GeV and 190 GeV.
Our results on the visible decays of the Higgs are based on the
study of efficiencies and backgrounds in the search for the Standard
Model Higgs described in reference \cite{Janot92}.  For the invisible
decays of the Higgs we considered only the channel  HZ with
$Z \ra e^+e^-$   or $Z \ra \mu^+\mu^-$, giving a signature of two leptons
plus missing transverse momentum. The requirement that the invariant
mass of the two leptons must be close to the Z mass can kill most of
the background from WW and $\gamma\gamma$ events; the background from
ZZ events with one of the Z decaying to neutrinos is small and the
measurement of the mass recoiling against the two leptons allows
to further reduce it, at least for $M_H$ not too close to $M_Z$.
Hadronic decays of the Z were not considered, since the background
from WW and $We\nu$ events is very large, and b-tagging is much less
useful than in the search for $Z H_{SM}$ with
$Z \ra  \nu \bar{\nu}$, since the $Zb\bar{b}$ branching ratio
is much smaller than $Hb\bar{b}$ in the standard model.
The dashed and dotted curves on figure 2 show
the exclusion contours in the $\epsilon^2$ vs. $M_H$ plane
that can be explored at LEPII, for the given centre-of-mass energies.
Again, these contours are valid irrespective of whether
the Higgs decays visibly, as in the standard model, or invisibly.
The possibility of invisible Higgs decay
is also very interesting from the point of
view of a linear $e^+ e^-$ collider at higher
energy \cite{EE500}.

{\bf Acknowledgements}\\
This work was supported by DGICYT and by
Accion Integrada Hispano-Portuguesa, under grant
numbers PB92-0084 and HP-50B. F. de Campos was
supported by a fellowship from the Brazilian
government (CNPq). He thanks the organizers,
especially C. A. Savoy for the hospitality at Moriond.

\section*{Figure Captions}
\noindent
\vskip .5cm
\centerline{
\protect{\hbox{\psfig{file=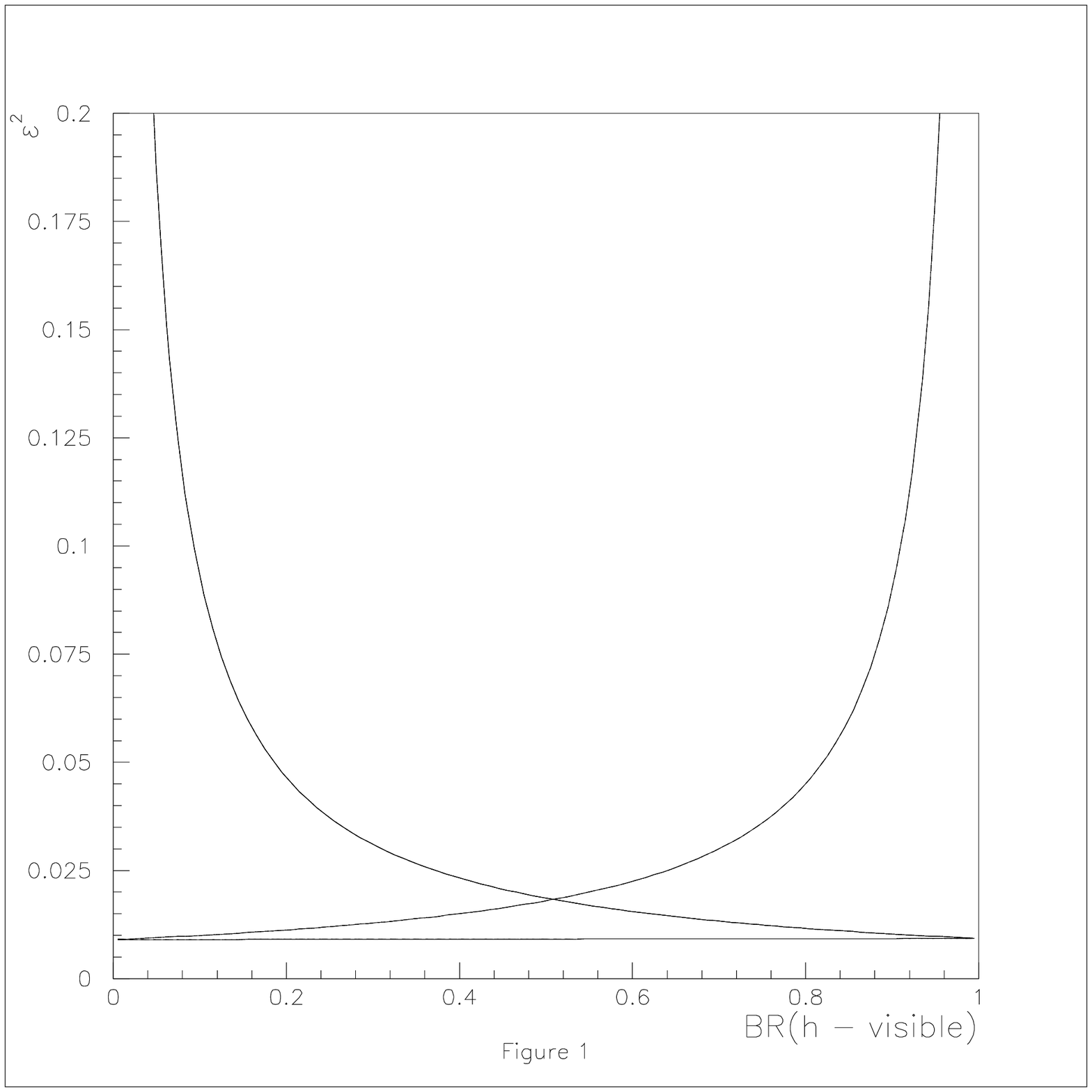,width=12cm}}}}
\vskip .3cm
{\footnotesize
Figure 1 shows the exclusion contours in the plane $\epsilon^2$
vs. $BR(H \ra $ visible) for the particular choice $m_H = 50$ GeV.
The two curves corresponding to the searches for visible (curve A)
and invisible (curve B) decays are combined to give the final
bound, which holds irrespective of the value of $BR(H \ra $ visible).}
\newline
\newpage
\noindent
\centerline{
\protect{\hbox{\psfig{file=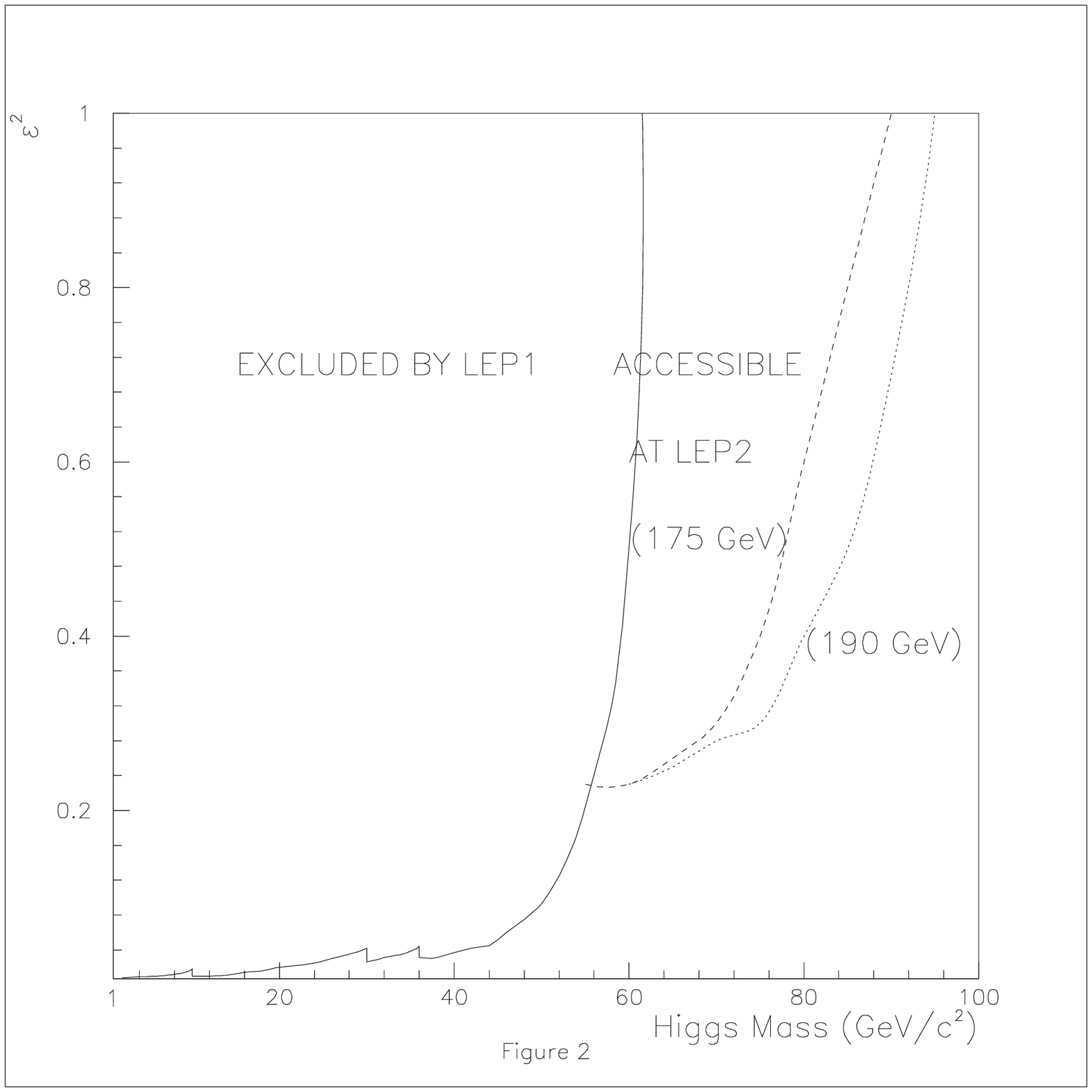,width=12cm}}}}
\vskip .5cm
{\footnotesize The solid curve shows the region in the $\epsilon^2$ vs. $m_H$
that can be excluded by the present LEP analyses. The dashed
and dotted curves on figure 2 show the exclusion contours in
the $\epsilon^2$ vs $m_H$ plane that can be explored at LEPII,
for the given centre-of-mass energies.}

\end{document}